\documentclass[aps,prd,onecolumn,nofootinbib,superscriptaddress]{revtex4-2}
\usepackage{amsmath,amssymb,bm}
\usepackage{booktabs}
\usepackage[colorlinks=true,linkcolor=blue,citecolor=blue,urlcolor=blue]{hyperref}
\newcommand{\eH}{\epsilon_H}

\begin{document}

\title{Isaacson energy--momentum tensor of high-frequency gravitational waves in general $f(R,G)$ gravity}

\author{Farzad Milani}
\email{fmilani@tvu.ac.ir}
\affiliation{Department of Basic Sciences, Technical and Vocational University (TVU), Tehran, Iran}

\date{\today}

\begin{abstract}
We construct the averaged energy--momentum tensor of high-frequency gravitational waves, in the sense of Isaacson, for a general function of the Ricci scalar and the Gauss--Bonnet invariant on de~Sitter, extending a recent $f(R)$ result. At linear order, every such theory behaves like $f(R)$ with a renormalised second derivative. At second order, tensor waves carry the Isaacson stress times the effective Planck mass, while scalar waves pick up an unsuppressed, traceless and non-null Gauss--Bonnet term. In the degenerate classes $F(R+\lambda G)$ and $R-2\Lambda+F(G)$, its size does not depend on $F$. For an isotropic ensemble, it lowers the energy that scalar waves contribute to the background by a factor between $1/3$ and $1$; for a single beam, the averaged energy violates the dominant energy condition. In wave packets, conservation fixes an accompanying gradient term. Both terms are gauge invariant at leading order, and the main result is confirmed in four ways: two independent implementations of the field equations, an exact special case and a direct variation of the action. Off de~Sitter, theories with a non-degenerate Hessian have no high-frequency regime. The degenerate class keeps the de~Sitter structure with ordinary slow-roll corrections. Still, stability of its scalar mode in the radiation and matter eras makes the Gauss--Bonnet correction negligible at late times.
\end{abstract}

\maketitle

\section{Introduction}\label{sec:intro}
Gravitational waves carry energy and momentum, and when their wavelength is short compared with the curvature radius of the background, this statement can be made precise. Averaging the quadratic part of the field equations over several wavelengths gives an effective, covariantly conserved source for the background. This is the construction of Isaacson~\cite{Isaacson1,Isaacson2}, which rests on the averaging procedure introduced by Brill and Hartle~\cite{BrillHartle} and was later recast in terms of an averaged Lagrangian by MacCallum and Taub~\cite{MacCallumTaub}. Textbook accounts can be found in Refs.~\cite{MTW,Maggiore}, and Ref.~\cite{FlanaganHughes} gives a compact review. The same idea, applied to long-wavelength cosmological perturbations, underlies the effective energy--momentum tensor of Abramo, Brandenberger and Mukhanov~\cite{ABM,MAB}; a rigorous formulation of the short-wavelength limit in cosmology is due to Green and Wald~\cite{GreenWald}.

In theories beyond general relativity, the effective stress tensor has been derived for several classes of models by Stein and Yunes~\cite{SteinYunes}, and it is the starting point of recent work on gravitational-wave memory in theories with extra fields~\cite{HYZ,Zosso}. Most of this work assumes an asymptotically flat background and massless fields. On a de~Sitter background, even the quadrupole formula of general relativity requires care~\cite{ABK,DateHoque1,DateHoque2}. For $f(R)$ gravity on de~Sitter, the averaged tensor was obtained recently by Tretyakov and Petrov (TP)~\cite{TP}, including the massive scalar mode that $f(R)$ adds to the two tensor polarisations~\cite{BerryGair,Liang}. In their result the scalar waves contribute a term $(6/f'_0)\langle\partial\phi^{(1)}\partial\phi^{(1)}\rangle$, which has the form of null dust.

The question we address is what happens when the Gauss--Bonnet invariant $G$ is allowed into the action through a general function $f(R,G)$. Gauss--Bonnet modifications of this type were proposed as alternatives to dark energy~\cite{NojiriOdintsov,Cognola,DeFeliceTsujikawaFG} and are covered in the reviews~\cite{SotiriouFaraoni,DeFeliceTsujikawaLRR,CapozzielloDeLaurentis,NOO}. Their perturbations are delicate. Already for actions quadratic in curvature, a massive spin-2 ghost appears unless the Weyl-squared combination is absent~\cite{Chiba,DHT}, and in Gauss--Bonnet cosmologies ghost-free conditions restrict the couplings~\cite{CdCDF}. For general $f(R,G)$, De Felice and Suyama showed that on non-de~Sitter cosmological backgrounds the short-wavelength scalar modes obey $\omega^2\propto k^4$ and are either unstable or superluminal~\cite{DFS}; similar matter instabilities were found in $f(G)$ models~\cite{DMT}. De Felice and Tanaka then showed that on anisotropic backgrounds a ghost is unavoidable unless the Hessian of $f$ is degenerate~\cite{DFT}, and De Felice, Gerard and Suyama worked out the cosmological perturbations with a perfect fluid~\cite{DGS}. Theories with a degenerate Hessian are equivalent to a single non-dynamical scalar coupled to $R$ and $G$, and therefore belong to the Horndeski class~\cite{Horndeski,KYY}, whose causal structure has been studied in Ref.~\cite{Reall}. On de~Sitter, however, the $k^4$ term drops out~\cite{DFS}. That makes de~Sitter the natural background on which to define the high-frequency energy--momentum tensor for \emph{general} $f(R,G)$, and it is where we start.

The outcome is simple to state. On de~Sitter the linear theory of any $f(R,G)$ is that of $f(R)$ with $f_{RR}$ replaced by $K\equiv f_{RR}+8H^2f_{RG}+16H^4f_{GG}$. At second order, the tensor waves give $f_R$ times the Isaacson tensor, while the scalar waves give the TP term plus a traceless, non-null piece proportional to $\Delta=8H^4f_R(f_{RG}+4H^2f_{GG})/K$. This piece vanishes in $f(R)$ but not in $R+F(G)$, nor in the ghost-free degenerate class. In the class $F(R+\lambda G)$, its size relative to the TP term depends only on $\lambda H^2$, whatever $F$ is, and in $R-2\Lambda+F(G)$ it removes two thirds of the scalar-wave energy. We have checked that it is gauge invariant at leading order and reproduced it by independent routes. For wave packets, conservation fixes a companion null term enhanced by $1/(HL_{\rm env})$, which is gauge invariant as well. As a source for the background, an isotropic ensemble of scalar waves behaves as radiation with a reduced energy density, while a single beam violates the dominant energy condition. Off de~Sitter only the degenerate class keeps a high-frequency regime; for $f=R+\alpha(R+\lambda G)^2$ the de~Sitter structure survives with $\mathcal O(\eH)$ corrections and nothing enhanced, and stability of the same class in the radiation and matter eras limits the Gauss--Bonnet correction today to a negligible size.

The derivations are given in the appendices. Every relation there was also checked with SymPy; the scripts and their outputs are deposited in Ref.~\cite{code}, and Appendix~\ref{app:numerics} collects the few statements that are purely numerical.

\section{Field equations}\label{sec:fieldeqs}
We take
\begin{equation}
S=\frac{1}{2\kappa^2}\int d^4x\sqrt{-g}\,f(R,G)+S_m ,\qquad G=R^2-4R_{\mu\nu}R^{\mu\nu}+R_{\mu\nu\rho\sigma}R^{\mu\nu\rho\sigma}.
\end{equation}
With $\phi\equiv f_R$, $\xi\equiv f_G$ and $V\equiv Rf_R+Gf_G-f$ the metric equations read
\begin{equation}
E_{\mu\nu}\equiv \phi\,G_{\mu\nu}+(g_{\mu\nu}\Box-\nabla_\mu\nabla_\nu)\phi+\tfrac12 g_{\mu\nu}V+4P_{\mu\alpha\nu\beta}\nabla^\alpha\nabla^\beta\xi=\kappa^2T_{\mu\nu},
\label{eq:E}
\end{equation}
where $P_{\mu\alpha\nu\beta}$ is the double dual of the Riemann tensor (Appendix~\ref{app:fieldeqs}). Equation~(\ref{eq:E}) holds for any $f$, degenerate Hessian included. On de~Sitter, $R_0=12H^2$ and $G_0=24H^4$, and Eq.~(\ref{eq:E}) requires
\begin{equation}
f_0=6H^2f_R+24H^4f_G .
\label{eq:dScond}
\end{equation}
There $P_{\mu\alpha\nu\beta}=H^2(g_{\mu\nu}g_{\alpha\beta}-g_{\mu\beta}g_{\alpha\nu})$. From now on all derivatives of $f$ are evaluated on the background.

We use three examples. The first,
\begin{equation}
f=R-2\Lambda+\gamma G^2 ,
\label{eq:stablemodel}
\end{equation}
has a de~Sitter solution with $\Lambda=3H^2-288\gamma H^8$, and, as we shall see, a stable one for $0<\gamma H^6<1/384$. We use it as the simplest case of a stable de~Sitter point with a Gauss--Bonnet term, not as a cosmological model: as late-time dark energy, models of the type $R+F(G)$ are excluded by an ultraviolet instability of matter perturbations~\cite{DMT}. It belongs to the class $R-2\Lambda+F(G)$, whose Hessian is degenerate because $f_{RR}=f_{RG}=0$. The other degenerate class, $f=F(R+\lambda G)$, contains the earlier Gauss--Bonnet extensions of the Starobinsky model; both classes are treated in Appendix~\ref{app:models}. Two further models illustrate what happens when the de~Sitter point is unstable: $f=R+\alpha R^2+\gamma G^2$ and $f=R+\alpha(R+\lambda G)^2$. For $\gamma=0$ or $\lambda=0$ (the Starobinsky model~\cite{Starobinsky}) Eq.~(\ref{eq:dScond}) has no solution, so in these two models the exact de~Sitter solution exists only because of the Gauss--Bonnet term (Appendix~\ref{app:linear}). The second of them is the member of the degenerate class with $F=X+\alpha X^2$; adding $-2\Lambda$ to $F$ makes its de~Sitter point stable as well (Appendix~\ref{app:models}).

\section{Linear theory on de Sitter}\label{sec:linear}
For an arbitrary metric perturbation on de~Sitter one has $\delta G=4H^2\delta R$. As a result the linearised Eq.~(\ref{eq:E}) sees the two scalars $\phi$ and $\xi$ only through the combination
\begin{equation}
\delta\Phi\equiv\delta\phi+4H^2\delta\xi=K\,\delta R,\qquad K\equiv f_{RR}+8H^2f_{RG}+16H^4f_{GG},
\end{equation}
and takes the form
\begin{equation}
E^{(1)}_{\mu\nu}=f_R\big[G^{(1)}_{\mu\nu}+3H^2h_{\mu\nu}\big]+(g_{\mu\nu}\Box-\nabla_\mu\nabla_\nu)\delta\Phi+3H^2g_{\mu\nu}\delta\Phi .
\label{eq:E1}
\end{equation}
The tensor sector is $f_R$ times that of general relativity with a cosmological constant, and there is a single scalar mode,
\begin{equation}
(\Box-m^2)\,\delta\Phi=0,\qquad m^2=\frac13\Big(\frac{f_R}{K}-12H^2\Big).
\label{eq:mass}
\end{equation}
For $f_G\to0$ this is the mass found by TP~\cite{TP}, and it agrees with the de~Sitter effective mass of Ref.~\cite{DFS} (their Eq.~(87)). The Hessian determinant $f_{RR}f_{GG}-f_{RG}^2$ plays no role here, in line with the absence of a $k^4$ term on de~Sitter~\cite{DFS}. For the model~(\ref{eq:stablemodel}) one finds $m^2=1/(96\gamma H^4)-4H^2$, which is positive for $\gamma H^6<1/384$; the scalar is moreover light, $m<3H/2$, for $1/600<\gamma H^6<1/384$ (Appendix~\ref{app:models}). In the two other models with $\alpha>0$ the scalar is tachyonic, and their de~Sitter point is an unstable fixed point (Appendix~\ref{app:linear}); we return to the meaning of this in Sec.~\ref{sec:offdS}.

\section{Second order and spatial averaging}\label{sec:second}
We use conformal time, $a=-1/(H\eta)$, and give all tensors in conformal components $(\eta,x,y,z)$. The waves move along $z$, $\langle\cdots\rangle$ is the exact spatial average over a wavelength, and we define $\kappa^2t_{\mu\nu}\equiv-\langle E^{(2)}_{\mu\nu}\rangle$.

\paragraph{Tensor waves.}
For the exact de~Sitter tensor mode the leading result is
\begin{equation}
\kappa^2t_{\mu\nu}=\frac{f_R}{4}\,\langle\partial_\mu h_{ab}\,\partial_\nu h^{ab}\rangle ,
\label{eq:tensor}
\end{equation}
that is, $f_R$ times the Isaacson tensor. The second derivatives $f_{RR}$, $f_{RG}$, $f_{GG}$ appear only at relative order $(H/k)^2$, and neither $\langle R^{(2)}\rangle$ nor $\langle G^{(2)}\rangle$ is enhanced by $k^2$ (Appendix~\ref{app:second}).

\paragraph{Scalar waves.}
The scalar wave is most easily described as a conformal perturbation, $g=a^2(1+2\sigma)\eta_{\mu\nu}$, with $\delta\Phi=-2f_R\sigma$; this solves the linear equations exactly if and only if $(\Box-m^2)\sigma=0$. For a single mode $\sigma={\rm Re}[s(\eta)e^{iqz}]$ we first use the mode equation to reduce the averaged source, without approximation, to a quadratic form in $s$ and $s'$, and only then insert the Hankel asymptotic series in $X\equiv-q\eta=k_{\rm phys}/H\gg1$ (Appendix~\ref{app:second}). The result is
\begin{equation}
\kappa^2 t_{\mu\nu}=6f_R\,\langle\partial_\mu\sigma\,\partial_\nu\sigma\rangle-\frac{2\Delta}{H^2}\,\langle\sigma_{,z}^2\rangle\,\big(\hat t\hat t+\hat x\hat x+\hat y\hat y-\hat z\hat z\big)_{\mu\nu}
\label{eq:main}
\end{equation}
with
\begin{equation}
\Delta=\frac{8H^4f_R\,(f_{RG}+4H^2f_{GG})}{K},
\label{eq:Delta}
\end{equation}
up to corrections of relative order $H/k$; $\hat t,\hat x,\hat y,\hat z$ denote the coordinate unit covectors.

For $f_{RG}=f_{GG}=0$ Eq.~(\ref{eq:main}) reduces to the scalar part of TP's Eq.~(123). The new $\Delta$ term is traceless and separately conserved, with radiation scaling, but it is not null: it shifts the energy density, $\kappa^2t_{\eta\eta}=(6f_R-2\Delta/H^2)\langle\sigma_{,z}^2\rangle$, and leaves the momentum flux $\kappa^2|t_{\eta z}|=6f_R\langle\sigma_{,z}^2\rangle$ untouched. Its size is set by $f_{RG}+4H^2f_{GG}=d\xi/dR|_{\rm dS}$, the response of the Gauss--Bonnet coupling to the wave, rather than by the Hessian determinant. Two cases are worth stating explicitly (Appendix~\ref{app:models}). In the whole class $R-2\Lambda+F(G)$ one has $f_{RG}=0$ and $\Delta=2H^2$, independently of $F$; for the model~(\ref{eq:stablemodel}) the energy density of a scalar wave is therefore one third of what the TP term alone would give. In the degenerate class $f=F(R+\lambda G)$,
\begin{equation}
\frac{\Delta}{3H^2f_R}=\frac{8u}{3(1+4u)},\qquad u=\lambda H^2,
\label{eq:DeltaClass}
\end{equation}
again for every $F$, so that the factor $1-\Delta/(3H^2f_R)=(3+4u)/(3+12u)$ decreases monotonically from $1$ at $u=0$ to $1/3$ as $u\to\infty$. It never changes sign.

\section{Checks}\label{sec:validation}
The $\Delta$ term is what remains after a cancellation among four-derivative structures of order $(k/H)^4$, so an error in any of them would show up as a spurious result of exactly this kind. We have therefore checked it in five independent ways. First, the assembled second-order field equations obey $\nabla^\mu E_{\mu\nu}=0$ identically, off shell, for an arbitrary profile $\sigma(\eta,z)$ (Appendix~\ref{app:bianchi}). Second, the gauge-dependent part of the averaged source is suppressed by $(H/k)^2$ relative to Eq.~(\ref{eq:main}) for scalar, time-like and longitudinal gauge vectors (Appendix~\ref{app:gauge}). Third, an implementation based on the closed-form curvature of a conformally flat metric reproduces the full off-shell $E^{(2)}$ identically; in that implementation the sign of the double dual is not assumed but fixed by the Bianchi identity (Appendix~\ref{app:conformal}). Fourth, for $f_R=18H^2K$ (Hankel index $1/2$) the averaged source is an exact polynomial in $X$ and agrees with Eq.~(\ref{eq:main}) without any asymptotic expansion (Appendix~\ref{app:conformal}). Finally, varying $\int\sqrt{-g}f$ directly reproduces Eq.~(\ref{eq:E}), through second order, for all components in which $\Delta$ appears (Appendix~\ref{app:action}).

All of these checks, and every relation in the appendices, were carried out in SymPy~\cite{code}. The scripts were recheched and improved with the help of an AI assistant (Claude, Anthropic); the author checked every result against the outputs of the deposited code.

\section{Wave packets and scale separation}\label{sec:packets}
A traceless term with $\rho\neq|{\rm flux}|$ raises an obvious question: how is energy transported? To answer it we take a narrow-band packet $\sigma={\rm Re}\int dp\,A(p)s_p(\eta)e^{ipz}$. Its carrier-averaged source is
\begin{equation}
\int\!\!\int dq_1dq_2\,A(q_1)A^*(q_2)\,\mathcal K(\eta;q_1,q_2)\,e^{i(q_1-q_2)z},
\end{equation}
where $\mathcal K$ is the exact two-mode bilinear form of $E^{(2)}$ (Appendix~\ref{app:packets}). To first order in envelope gradients,
\begin{align}
\kappa^2t_{\mu\nu}={}&6f_R\langle\sigma_{,z}^2\rangle\hat k_\mu\hat k_\nu\nonumber\\
&-\frac{2\Delta}{H^2}\Big[\langle\sigma_{,z}^2\rangle\big(\hat t\hat t+\hat x\hat x+\hat y\hat y-\hat z\hat z\big)_{\mu\nu}+\eta\,\partial_u\langle\sigma_{,z}^2\rangle\,\hat k_\mu\hat k_\nu\Big],
\label{eq:packet}
\end{align}
with $\hat k=\hat t-\hat z$, $u=z-\eta$ and $\eta\partial_u=-H^{-1}\partial/\partial\ell_{\rm phys}$. The coefficient of the gradient term is not a free parameter; conservation fixes it. The gradient term is also gauge invariant at leading order. For the three families of gauge vectors used for the single mode, the change of the two-mode kernel at first order in the envelope gradient grows at most like $X$, whereas the term itself grows like $X^3$ (Appendix~\ref{app:gauge}).

The gradient term integrates to zero over the packet. The total energy is therefore proportional to $6f_R-2\Delta/H^2$ and the total momentum to $6f_R$, so the inequality $E\neq P$ holds for the packet as a whole and not only for a plane wave. Locally the picture is different. Relative to the monopole the gradient term is of order $1/(HL_{\rm env})$, which is large for any sub-Hubble envelope, and positivity of the local energy density requires $HL_{\rm env}\gtrsim|\Delta|/(3H^2f_R)$. In the Gauss--Bonnet sector, then, scale separation is controlled by the envelope length in Hubble units, not by the wavelength alone. For a statistically homogeneous and isotropic ensemble the gradient term averages out; we return to such ensembles in Sec.~\ref{sec:backreaction}.

\section{Backreaction}\label{sec:backreaction}
Averaged over many wavelengths, the waves enter the background equations as an additional source, $E^{(0)}_{\mu\nu}=\kappa^2(T_{\mu\nu}+t_{\mu\nu})$. The tensor waves contribute $f_R$ times the Isaacson tensor and need no further comment. For the scalar waves it is convenient to write Eq.~(\ref{eq:main}) in a local orthonormal frame, for a plane wave travelling along a unit vector $\hat n$, with $F\equiv\langle(\hat n\cdot\partial\sigma)^2\rangle$ and $\tilde\Delta\equiv\Delta/H^2$:
\begin{gather}
\kappa^2\rho=(6f_R-2\tilde\Delta)F,\qquad \kappa^2S_i=6f_RF\,\hat n_i,\nonumber\\
\kappa^2p_{ij}=6f_RF\,\hat n_i\hat n_j-2\tilde\Delta F\,(\delta_{ij}-2\hat n_i\hat n_j),
\label{eq:beam}
\end{gather}
where $S_i$ is the energy flux. The trace vanishes, as it must.

For a statistically homogeneous and isotropic ensemble, $\langle\hat n_i\hat n_j\rangle=\delta_{ij}/3$, the flux averages out and $p_{ij}=\tfrac13\rho\,\delta_{ij}$. The scalar waves then act on the background exactly as radiation does, but with energy density
\begin{equation}
\kappa^2\rho=6f_R\Big(1-\frac{\Delta}{3H^2f_R}\Big)F ,
\label{eq:rhoiso}
\end{equation}
instead of $6f_RF$. The factor in brackets is the whole effect of the Gauss--Bonnet coupling on the backreaction of scalar waves on de~Sitter. It equals $1/3$ in the class $R-2\Lambda+F(G)$, and $(3+4u)/(3+12u)$ in the class $F(R+\lambda G)$, so it lies between $1/3$ and $1$ in all the models considered here. Gauss--Bonnet terms therefore reduce the gravitating energy of a given scalar amplitude; they never enhance it, and they never make it negative.

A single beam is less tame. From Eq.~(\ref{eq:beam}), $\kappa^2(\rho-|S|)=-2\tilde\Delta F$, so for $\Delta>0$ the energy flux exceeds the energy density and the dominant energy condition fails. An observer moving along the beam with speed $v$ measures
\begin{equation}
\kappa^2\rho'=\frac{6f_R(1-v)^2-2\tilde\Delta(1-v^2)}{1-v^2}\,F ,
\end{equation}
which is negative for $v>(3f_R-\tilde\Delta)/(3f_R+\tilde\Delta)$. The null energy condition holds along the beam, where $t_{\mu\nu}k^\mu k^\nu=0$, but along a null direction transverse to it $t_{\mu\nu}k^\mu k^\nu=(6f_R-4\tilde\Delta)F/\kappa^2$, which is negative when $\Delta/(3H^2f_R)>1/2$. This happens in the class $R-2\Lambda+F(G)$ and in $F(R+\lambda G)$ for $u>3/4$. An effective stress tensor built from the gravitational sector is not obliged to satisfy energy conditions, and in $f(R)$ the scalar term, being null dust, satisfies them only marginally. Here the violation is carried by the averaged energy of the waves themselves; it disappears once the beams are summed over directions.

How large can the effect be in practice? The result holds on de~Sitter and, for the degenerate class, along a slow departure from it (Sec.~\ref{sec:offdS}); it does not describe the radiation or matter eras. A late-time application would need a stable de~Sitter attractor, and the stability of the scalar mode in the earlier eras then puts a strong limit on $u$, which we derive in Sec.~\ref{sec:offdS}: for the class $F(R+\lambda G)$ the Gauss--Bonnet correction today is below $3\times10^{-11}$. During inflation the correction can be larger: if the radiation era starts at a Hubble rate close to the inflationary one, the same bound allows $u$ up to about $3/20$, that is, a reduction of the energy by up to about 25 per cent. There, however, the sub-horizon scalar modes are quantum fluctuations rather than classical waves, and we do not know of a mechanism that would produce a classical scalar-wave background large enough to matter. We therefore regard Eq.~(\ref{eq:rhoiso}) as a structural result: it fixes how the energy of scalar waves enters the background equations in these theories, whatever produces the waves.

\section{Off de Sitter}\label{sec:offdS}
\paragraph{Non-degenerate class.}
On spatially flat FLRW backgrounds the short-wavelength scalar modes satisfy $\omega^2=B_1k^2/a^2+B_2k^4/a^4$~\cite{DFS}, and to leading slow-roll order
\begin{equation}
B_2\simeq-\frac{64\,\dot H^2\,(f_{RR}f_{GG}-f_{RG}^2)}{3Kf_R}.
\label{eq:B2}
\end{equation}
For $f=R+\alpha R^2+\gamma G^2$ with $\alpha,\gamma>0$ this is negative, and the modes grow at the rate $\sqrt{|B_2|}\,k_{\rm phys}^2$. In the parameter range of Appendix~\ref{app:numerics} this rate exceeds $H$ once $k_{\rm phys}\gtrsim(20$--$120)H$. For $\gamma<0$ and $\alpha>16|\gamma|H^4$, so that $K>0$, $B_2$ is positive instead; the modes are then dispersive, with group velocity above unity for $k_{\rm phys}\gtrsim H/\eH$ (Appendix~\ref{app:numerics}). Either way there is no high-frequency regime. None of this is new, since it follows directly from Ref.~\cite{DFS}; we include it because it singles out the degenerate class as the only one in which our construction can be continued away from de~Sitter.

\paragraph{Degenerate class: $f=R+\alpha(R+\lambda G)^2$.}
Here $B_2\equiv0$. To leading slow-roll order the background equation gives
\begin{equation}
\eH=\frac{D}{18H^2K},\qquad D=1-48\alpha H^2u(1+2u),\qquad u=\lambda H^2,
\end{equation}
which reduces to the Starobinsky result~\cite{Starobinsky} for $\lambda\to0$; the derivation is in Appendix~\ref{app:offdS}. Two regimes should be distinguished. When $u\lesssim\eH$ the model is Starobinsky-like and $\Delta/(3H^2f_R)\simeq\frac83u$ is itself of order $\eH$. When $u\gg\eH$ the background leaves the Gauss--Bonnet-supported de~Sitter point along the growing homogeneous scalar mode, $H=H_{\rm dS}(1+\delta e^{st})$ with $s^2+3s+m^2=0$.

In both regimes the scalar sound speed can be obtained analytically. In slow roll
\begin{equation}
c_s^2=1-\frac{8\eH u\,(24u^2+16u+3)}{3(2u+1)^2(4u+1)},
\end{equation}
and the tensor speed read off from our linear equations coincides, on an arbitrary FLRW background, with $c_T^2=(F+4\ddot\xi)/(F+4H\dot\xi)$ of Ref.~\cite{DGS} (Appendix~\ref{app:offdS}). The deviation $c_T^2-1$ is of order $\eH$ and is tied to the departure from the high-curvature de~Sitter point, so the late-time bound from GW170817 does not apply to it directly; a version of the model meant to describe late-time acceleration would have to satisfy that bound~\cite{GW170817,EZ,CV}.

What does a high-frequency stress tensor mean on an unstable background? The relevant comparison is between time scales. The background leaves the de~Sitter point on the time scale $1/s$, with $s$ between $0.03\,H$ and $0.7\,H$ for $1/200\le u\le1$ (Appendix~\ref{app:numerics}), whereas the waves oscillate on the time scale $1/k_{\rm phys}$ with $k_{\rm phys}\gg H$. The averaging that defines $t_{\mu\nu}$ needs only this separation, exactly as for gravitational waves during inflation, which is itself a slow departure from a de~Sitter phase. The next paragraph shows that the separation is not spoilt: along the growing mode the result keeps the de~Sitter form with the local Hubble rate. Stability of the de~Sitter point, when one wants it, is a property of the model and not of the result, as the model~(\ref{eq:stablemodel}) shows.

On the growing-mode background we evaluated the averaged second-order source to $\mathcal O(\eH)$ for eight values of $u$ between $1/200$ and $1$ (Appendix~\ref{app:numerics}). In every case the terms of order $q^4$ and $q^3$ vanish identically, and the $q^2$ term equals Eq.~(\ref{eq:main}) evaluated with the local Hubble rate, plus remainders proportional to $\eH$. No enhanced terms appear, and the high-frequency hierarchy is preserved. The size of the remainders does depend on $u$. Relative to the de~Sitter value they stay between $0.3\,\eH$ and $1.5\,\eH$ for $u\le1/20$, but they grow with $u$ and reach $4$ to $11$ times $\eH$ at $u=1$. For large $u$ the de~Sitter form is therefore accurate only when $\eH$ is correspondingly small. We did not obtain the remainders in closed form as functions of $u$; the symbolic expressions grow too quickly, so we give them on this grid instead.

\paragraph{Radiation and matter eras.}
The sound speed of the degenerate class was derived on an arbitrary FLRW background, so it also applies when the expansion is driven by a perfect fluid; the fluid perturbations enter the scalar equations only at order $q^0$ and do not change the wave speed. For a fluid with constant equation of state $w$, so that $\dot H=-\tfrac32(1+w)H^2$, and with $\alpha H^2\ll1$, we find
\begin{equation}
c_s^2=1-\frac{8u(1+w)}{1+4u},\qquad u=\lambda H^2
\label{eq:csfluid}
\end{equation}
(Appendix~\ref{app:offdS}). For $u\to\infty$ this becomes $-1-2w$, the value found for $R+F(G)$ in Ref.~\cite{DMT}, and the same expression follows from Eq.~(33) of Ref.~\cite{DGS} with $\dot F/(4H^2\dot\xi)=1/(4u)$. The mode is stable only for $u<1/4$ in the matter era and $u<3/20$ in the radiation era. Since $u=\lambda H^2$ grows towards the past, stability since matter--radiation equality already requires $u_0<\tfrac14(H_0/H_{\rm eq})^2\approx10^{-11}$ today, and therefore $\Delta/(3H_0^2f_R)<3\times10^{-11}$. The same parameter $u$ that sets the size of the Gauss--Bonnet term in the stress tensor also controls the stability of the theory in the eras before it, and the second requirement makes the first effect negligible at late times.

\section{Discussion}\label{sec:limits}
On de~Sitter the Isaacson tensor of a general $f(R,G)$ theory is well defined. Compared with $f(R)$ there is one genuinely new ingredient besides the replacement $f_{RR}\to K$: the non-null scalar term proportional to $\Delta$, with the gradient term that conservation attaches to it in a packet. Its size is fixed by the response of the Gauss--Bonnet coupling to the wave. In the two degenerate classes it does not depend on the detailed form of $f$: it is $2/3$ of the TP term in $R-2\Lambda+F(G)$ and $8u/[3(1+4u)]$ of it in $F(R+\lambda G)$. For an isotropic ensemble this simply lowers the energy that scalar waves contribute to the background, by a factor between $1/3$ and $1$. For a single beam it is more striking, because the averaged energy then violates the dominant energy condition.

Whether the de~Sitter point is stable is a property of the model and not of the result. Away from de~Sitter the non-degenerate theories lose the high-frequency regime altogether, whereas the degenerate class keeps the de~Sitter structure, with corrections of order $\eH$.

The limitations are the following. The off-de~Sitter remainders are known on a grid of $u$ rather than in closed form, and only to first order in $\eH$. The packet analysis is one-dimensional; envelopes that vary across the direction of propagation have not been treated. The stability bound in the radiation and matter eras assumes a background close to general relativity, $\alpha H^2\ll1$. Finally, we have not found a setting in which classical scalar waves of appreciable amplitude live on a nearly de~Sitter background: at late times the stability bound makes the effect negligible, and during inflation the relevant sub-horizon modes are quantum. The results should therefore be read as a statement about how scalar gravitational waves gravitate in these theories, not as a prediction for a particular signal.

\section*{Declarations}
\paragraph{Funding.} No funding was received for this work.
\paragraph{Competing interests.} The author has no relevant financial or non-financial interests to disclose.
\paragraph{Data availability statement.} {\sloppy This manuscript has no associated data. The SymPy scripts behind the symbolic and numerical results, together with their reference outputs, are openly available on Zenodo~\cite{code}, \href{https://doi.org/10.5281/zenodo.23811207}{doi:10.5281/zenodo.23811207}.\par}

\appendix

\section{Field equations and the de Sitter background}\label{app:fieldeqs}
The double dual is $P_{\mu\alpha\nu\beta}=\tfrac14\epsilon_{\mu\alpha\rho\sigma}\epsilon_{\nu\beta\kappa\lambda}R^{\rho\sigma\kappa\lambda}$. In four dimensions
\begin{align}
P_{\mu\alpha\nu\beta}=s_P\Big[&-R_{\mu\alpha\nu\beta}+g_{\mu\nu}R_{\alpha\beta}-g_{\mu\beta}R_{\alpha\nu}-g_{\alpha\nu}R_{\mu\beta}+g_{\alpha\beta}R_{\mu\nu}\nonumber\\
&-\frac{R}{2}\big(g_{\mu\nu}g_{\alpha\beta}-g_{\mu\beta}g_{\alpha\nu}\big)\Big],\qquad s_P=-1 ,
\label{eq:P}
\end{align}
and $s_P=-1$ is the only value for which Eq.~(\ref{eq:E}) satisfies the Bianchi identity (Appendix~\ref{app:conformal}). $P$ is divergence free, and on de~Sitter $P_{\mu\alpha\nu\beta}=H^2(g_{\mu\nu}g_{\alpha\beta}-g_{\mu\beta}g_{\alpha\nu})$.

On de~Sitter $\phi$ and $\xi$ are constant, $G_{\mu\nu}=-3H^2g_{\mu\nu}$, and Eq.~(\ref{eq:E}) reduces to $E^{(0)}_{\mu\nu}=g_{\mu\nu}\big(-3H^2f_R+\tfrac12V_0\big)$. It vanishes for $V_0=6H^2f_R$. Since $V_0=12H^2f_R+24H^4f_G-f_0$, this is
\begin{equation}
f_0=6H^2f_R+24H^4f_G ,
\end{equation}
which is Eq.~(\ref{eq:dScond}).

\section{Linear theory on de Sitter}\label{app:linear}
\paragraph{Gauss--Bonnet scalar.}
On de~Sitter $R_{abcd}=H^2(g_{ac}g_{bd}-g_{ad}g_{bc})$, $R_{ab}=3H^2g_{ab}$ and $R=12H^2$. With $\delta g_{ab}=h_{ab}$, $h=g^{ab}h_{ab}$ and $\delta R=-3H^2h+g^{ab}\delta R_{ab}$,
\begin{align}
\delta(R^2)&=24H^2\,\delta R,\\
\delta(R_{ab}R^{ab})&=6H^2g^{ab}\delta R_{ab}-18H^4h=6H^2\,\delta R,\\
\delta(R_{abcd}R^{abcd})&=4H^2g^{bd}\delta R_{bd}+6H^4h-18H^4h=4H^2\,\delta R ,
\end{align}
so that $\delta G=(24-24+4)H^2\delta R=4H^2\delta R$ for arbitrary $h_{ab}$.

\paragraph{Linear field equations.}
On de~Sitter $\xi$ is constant, so the variation of $P$ does not contribute. We treat $\phi=f_R+\delta\phi$, $\xi=f_G+\delta\xi$ and $V=V_0+12H^2\delta\phi+24H^4\delta\xi$ as independent of $h_{\mu\nu}$ and use $4P^{(0)}_{\mu\alpha\nu\beta}\nabla^\alpha\nabla^\beta\delta\xi=4H^2(g_{\mu\nu}\Box-\nabla_\mu\nabla_\nu)\delta\xi$. All ten components of the linearised Eq.~(\ref{eq:E}) then take the form
\begin{equation}
E^{(1)}_{\mu\nu}=f_R\big[G^{(1)}_{\mu\nu}+3H^2h_{\mu\nu}\big]+(g_{\mu\nu}\Box-\nabla_\mu\nabla_\nu)\delta\Phi+3H^2g_{\mu\nu}\delta\Phi ,
\end{equation}
with $\delta\Phi=\delta\phi+4H^2\delta\xi$, which is Eq.~(\ref{eq:E1}). Inserting $\delta\phi=f_{RR}\delta R+f_{RG}\delta G$ and $\delta\xi=f_{RG}\delta R+f_{GG}\delta G$ with $\delta G=4H^2\delta R$,
\begin{equation}
\delta\Phi=\big(f_{RR}+4H^2f_{RG}\big)\delta R+4H^2\big(f_{RG}+4H^2f_{GG}\big)\delta R=K\,\delta R .
\end{equation}
If the Gauss--Bonnet term is dropped, that is, if $\delta\Phi$ is replaced by $\delta\phi$, then $E^{(1)}_{xy}$ is left with a residual $-4H^2\partial_x\partial_y\delta\xi$.

\paragraph{Trace and mass.}
On shell,
\begin{equation}
g^{\mu\nu}E^{(1)}_{\mu\nu}-h^{\mu\nu}E^{(0)}_{\mu\nu}=-f_R\,\delta R+3\Box\delta\Phi+12H^2\delta\Phi .
\end{equation}
In vacuum the left-hand side vanishes. With $\delta R=\delta\Phi/K$ this becomes $3\Box\delta\Phi-(f_R/K-12H^2)\delta\Phi=0$, that is
\begin{equation}
(\Box-m^2)\,\delta\Phi=0,\qquad m^2=\frac13\Big(\frac{f_R}{K}-12H^2\Big),
\end{equation}
which is Eq.~(\ref{eq:mass}). For $f_{RG}=f_{GG}=0$ it equals the mass of Ref.~\cite{TP}, $m^2=\tfrac23\big(f'/(2f'')-6H^2\big)$ in their notation.

\paragraph{Conformal scalar wave.}
For $g=a^2(1+2\sigma)\eta_{\mu\nu}$ one has $\delta R=-24H^2\sigma-6\Box\sigma$. On shell, $\Box\sigma=m^2\sigma$, so $\delta R=-2f_R\sigma/K$ and $\delta\Phi=-2f_R\sigma$.

\paragraph{Models.}
For $f=R+\alpha R^2+\gamma G^2$ the de~Sitter condition is $f_0-6H^2f_R-24H^4f_G=-6H^2(96H^6\gamma-1)=0$, and
\begin{gather}
K=2(\alpha+16H^4\gamma),\qquad m^2=-\frac{384H^6\gamma-1}{6(\alpha+16H^4\gamma)},\nonumber\\
f_{RG}+4H^2f_{GG}=8H^2\gamma,\qquad f_{RR}f_{GG}-f_{RG}^2=4\alpha\gamma .
\end{gather}
For $f=R+\alpha(R+\lambda G)^2$ the condition is $-6H^2(96H^6\alpha\lambda^2+48H^4\alpha\lambda-1)=0$, and
\begin{gather}
K=2\alpha(1+4\lambda H^2)^2,\qquad m^2=-\frac{384H^6\alpha\lambda^2+144H^4\alpha\lambda-1}{6\alpha(1+4\lambda H^2)^2},\nonumber\\
f_{RG}+4H^2f_{GG}=2\alpha\lambda(1+4\lambda H^2),
\end{gather}
with vanishing Hessian determinant. For $\gamma=\lambda=0$ the de~Sitter condition becomes $-6H^2=0$.

At the de~Sitter point the conditions fix $\gamma=1/(96H^6)$ in the first model and $\alpha=1/[48H^2u(1+2u)]$, with $u=\lambda H^2$, in the second. Substituting,
\begin{align}
f=R+\alpha R^2+\gamma G^2:&\qquad m^2=-\frac{3H^2}{1+6\alpha H^2},\qquad \frac{\Delta}{3H^2f_R}=\frac{2}{3(1+6\alpha H^2)},\\
f=R+\alpha(R+\lambda G)^2:&\qquad m^2=-\frac{16H^2u(1+3u)}{(1+4u)^2},\qquad \frac{\Delta}{3H^2f_R}=\frac{8u}{3(1+4u)} .
\end{align}
The scalar is therefore tachyonic in the first model for $\alpha>0$ and in the second for every $u>0$ (where $\alpha>0$ automatically), so the de~Sitter point is unstable.

\section{Stable de Sitter points and two degenerate classes}\label{app:models}
\paragraph{The class $R-2\Lambda+F(G)$.}
Here $f_R=1$, $f_{RR}=f_{RG}=0$ and $f_{GG}=F''(G_0)$, so the Hessian is degenerate, $K=16H^4F''$ and, from Eqs.~(\ref{eq:mass}) and~(\ref{eq:Delta}),
\begin{equation}
m^2=\frac{1}{48H^4F''}-4H^2,\qquad \Delta=\frac{8H^4\cdot4H^2F''}{16H^4F''}=2H^2,\qquad \frac{\Delta}{3H^2f_R}=\frac23 .
\end{equation}
The scalar is stable for $0<H^6F''<1/192$. For $F=\gamma G^2$ the field equations on de~Sitter, solved directly, give
\begin{equation}
\Lambda=3H^2-288\gamma H^8 ,
\end{equation}
which is positive for $\gamma H^6<1/96$. With $K=32\gamma H^4$,
\begin{equation}
m^2=\frac{1}{96\gamma H^4}-4H^2 ,
\end{equation}
so that $m^2>0$ for $\gamma H^6<1/384$ and $0<m^2<9H^2/4$ for $1/600<\gamma H^6<1/384$. The energy factor $1-\Delta/(3H^2f_R)$ equals $1/3$ for every $\gamma$.

For a heavy scalar the high-frequency expansion needs more than $k_{\rm phys}\gg H$. The first term of the series~(\ref{eq:hankel}) is $a_1/X=(1-m^2/2H^2)\,H/k_{\rm phys}$, so for $m\gg H$ the expansion requires $k_{\rm phys}\gg m^2/(2H)$.

\paragraph{The class $F(R+\lambda G)$.}
With $X=R+\lambda G$, $f_R=F'$, $f_{RR}=F''$, $f_{RG}=\lambda F''$ and $f_{GG}=\lambda^2F''$, so $f_{RR}f_{GG}-f_{RG}^2=0$ and $K=F''(1+4u)^2$ with $u=\lambda H^2$. Then
\begin{gather}
\frac{\Delta}{3H^2f_R}=\frac{8H^4F'\cdot\lambda F''(1+4u)}{3H^2F'\,F''(1+4u)^2}=\frac{8u}{3(1+4u)},\nonumber\\
1-\frac{\Delta}{3H^2f_R}=\frac{3+4u}{3(1+4u)},
\end{gather}
for every $F$. The second expression decreases monotonically from $1$ at $u=0$ to $1/3$ as $u\to\infty$; its derivative is $-8/[3(1+4u)^2]$. The mass is
\begin{equation}
m^2=\frac{F'}{3F''(1+4u)^2}-4H^2 ,
\end{equation}
so the de~Sitter point is stable when $F'/F''>12H^2(1+4u)^2$.

For $F=X-2\Lambda+\alpha X^2$, the condition~(\ref{eq:dScond}) gives
\begin{equation}
\Lambda=3H^2-144\,\alpha H^4u(1+2u),
\end{equation}
and
\begin{equation}
\frac{m^2}{H^2}=\frac{1}{6\alpha H^2(1+4u)^2}-\frac{8u(3+8u)}{(1+4u)^2}.
\end{equation}
The de~Sitter point is therefore stable for $\alpha H^2<1/[48u(3+8u)]$, a range in which $\Lambda>0$ automatically. Setting $\Lambda=0$ gives $\alpha H^2=1/[48u(1+2u)]$ and $m^2=-16H^2u(1+3u)/(1+4u)^2<0$, which is the model $R+\alpha(R+\lambda G)^2$ of Appendix~\ref{app:linear}. Three stable examples are $(u,\alpha H^2)=(1/50,1/10)$, $(1/20,1/50)$ and $(1,1/600)$, with $m^2/H^2=2177/2187$, $523/108$ and $12/25$ and energy factors $77/81$, $8/9$ and $7/15$.

\section{Second order on de Sitter}\label{app:second}
\paragraph{Tensor mode.}
The mode $h_{xy}=a^2A\,{\rm Re}[(1+iq\eta)e^{iq(z-\eta)}]$ solves the linear equations exactly. The averaged second-order equations are, without approximation,
\begin{align}
\langle E^{(2)}_{\eta\eta}\rangle&=-\tfrac18A^2q^2\big(126H^2K+2\eta^2q^2f_R-7f_R\big),\nonumber\\
\langle E^{(2)}_{\eta z}\rangle&=\tfrac14A^2\eta^2q^4f_R,\nonumber\\
\langle E^{(2)}_{zz}\rangle&=-\tfrac18A^2q^2\big(-42H^2K+2\eta^2q^2f_R+5f_R\big),\nonumber\\
\langle E^{(2)}_{xx}\rangle&=\langle E^{(2)}_{yy}\rangle=\tfrac18A^2q^2\big(42H^2K-f_R\big),
\end{align}
and $\langle E^{(2)}_{xy}\rangle=0$. With $h^{xy}=a^{-4}h_{xy}$ the same mode gives
\begin{align}
\langle\partial_\mu h_{ab}\,\partial_\nu h^{ab}\rangle&=2a^{-4}\langle\partial_\mu h_{xy}\,\partial_\nu h_{xy}\rangle\nonumber\\
&=A^2\eta^2q^4\,\hat k_\mu\hat k_\nu+\mathcal O(q^3),\qquad \hat k=\hat t-\hat z .
\end{align}
The leading terms of the $(\eta\eta)$, $(\eta z)$ and $(zz)$ components of $\langle E^{(2)}\rangle$ are $\mp\tfrac14f_RA^2\eta^2q^4$, with signs $-,+,-$, so that
\begin{equation}
\kappa^2t_{\mu\nu}=-\langle E^{(2)}_{\mu\nu}\rangle=\frac{f_R}{4}\,\langle\partial_\mu h_{ab}\,\partial_\nu h^{ab}\rangle\,\big[1+\mathcal O(H^2/k^2)\big],
\end{equation}
which is Eq.~(\ref{eq:tensor}), $f_R$ times the Isaacson tensor. Since $k_{\rm phys}=-qH\eta$, the remaining terms are of relative order $(H/k_{\rm phys})^2$, multiplied by $H^2K/f_R$ for those containing $K$. For the curvature scalars, $\langle R^{(2)}\rangle=\tfrac74A^2H^2q^2\eta^2$ and $\langle G^{(2)}\rangle=4H^2\langle R^{(2)}\rangle$.

\paragraph{Scalar mode equation and quadratic form.}
In conformal time $(\Box-m^2)\sigma=0$ reads
\begin{equation}
s''=\frac2\eta\,s'-\Big(q^2+\frac{m^2}{H^2\eta^2}\Big)s .
\end{equation}
Eliminating all second and higher $\eta$-derivatives with this equation reduces each component of $\langle E^{(2)}_{\mu\nu}\rangle$ exactly to
\begin{equation}
\mathcal A\,P+\mathcal B\,M+\mathcal C\,Q+\mathcal D\,W,\qquad P=|s|^2,\ M={\rm Re}(s^*s'),\ Q=|s'|^2,\ W={\rm Im}(s^*s'),
\end{equation}
with explicit coefficients $\mathcal A,\dots,\mathcal D$ depending on $\eta$; no other structure occurs.

\paragraph{Asymptotics.}
We use the mode
\begin{equation}
s=\frac Hq\,X e^{iX}\sum_{k=0}^{4}i^ka_kX^{-k},\qquad a_0=1,\quad a_k=a_{k-1}\frac{4\nu^2-(2k-1)^2}{8k},
\label{eq:hankel}
\end{equation}
with $\nu^2=9/4-m^2/H^2$ and $X=-q\eta$. The leading coefficients are
\begin{align}
\langle E^{(2)}_{\eta\eta}\rangle&=-\frac{H^2f_R(16H^4f_{GG}+16H^2f_{RG}+3f_{RR})}{K}X^2,&
\langle E^{(2)}_{\eta z}\rangle&=3H^2f_RX^2,\nonumber\\
\langle E^{(2)}_{zz}\rangle&=-\frac{H^2f_R(80H^4f_{GG}+32H^2f_{RG}+3f_{RR})}{K}X^2,&
\langle E^{(2)}_{xx}\rangle&=\langle E^{(2)}_{yy}\rangle=\Delta\,X^2 ,
\end{align}
where $\Delta$ is given by Eq.~(\ref{eq:Delta}). Using $K=f_{RR}+8H^2f_{RG}+16H^4f_{GG}$, the four coefficients combine into
\begin{equation}
-\langle E^{(2)}_{\mu\nu}\rangle=X^2\Big[3H^2f_R\,\hat k_\mu\hat k_\nu-\Delta\big(\hat t\hat t+\hat x\hat x+\hat y\hat y-\hat z\hat z\big)_{\mu\nu}\Big]+\mathcal O(X),\qquad \hat k=\hat t-\hat z ;
\end{equation}
for instance $-\langle E^{(2)}_{\eta\eta}\rangle X^{-2}=H^2f_R(3K-32H^4f_{GG}-8H^2f_{RG})/K=3H^2f_R-\Delta$. At the same order the mode gives $\langle\sigma_{,z}^2\rangle=\tfrac12q^2|s|^2=\tfrac12H^2X^2$ and, because the phase depends on $z-\eta$, $\langle\partial_\mu\sigma\,\partial_\nu\sigma\rangle=\langle\sigma_{,z}^2\rangle\,\hat k_\mu\hat k_\nu$. Replacing $X^2$ by $2\langle\sigma_{,z}^2\rangle/H^2$ we obtain
\begin{equation}
\kappa^2 t_{\mu\nu}=-\langle E^{(2)}_{\mu\nu}\rangle=6f_R\,\langle\partial_\mu\sigma\,\partial_\nu\sigma\rangle-\frac{2\Delta}{H^2}\,\langle\sigma_{,z}^2\rangle\,\big(\hat t\hat t+\hat x\hat x+\hat y\hat y-\hat z\hat z\big)_{\mu\nu}\,\big[1+\mathcal O(H/k)\big],
\end{equation}
which is Eq.~(\ref{eq:main}). The $\Delta$ part is traceless and satisfies $\rho=\sum p$, as radiation scaling requires.

\section{Bianchi identity}\label{app:bianchi}
For $g=a^2(1+2\epsilon\sigma)\eta_{\mu\nu}$ with arbitrary $\sigma(\eta,z)$, and with $\phi$, $\xi$, $V$ expanded to second order in $\delta R$, $\delta G$ (third derivatives of $f$ retained), $\nabla^\mu E_{\mu\nu}=0$ holds identically at orders $\epsilon^0$, $\epsilon^1$ and $\epsilon^2$ for $\nu=\eta,z$, without use of any field equation.

\section{Gauge invariance}\label{app:gauge}
Because $E_{\mu\nu}$ is a covariant functional of the metric, a first-order gauge transformation generated by $\zeta$ gives, on shell and exactly,
\begin{equation}
E^{(2)}[h+\mathcal L_\zeta\bar g]=E^{(2)}[h]-E^{(1)}[X_\zeta],\qquad X_\zeta=\tfrac12\mathcal L_\zeta\mathcal L_\zeta\bar g+\mathcal L_\zeta h .
\end{equation}
Averaging commutes with $E^{(1)}$, so the averaged source changes by $-E^{(1)}[\langle X_\zeta\rangle(\eta)]$, with $E^{(1)}$ from Eq.~(\ref{eq:E1}). We take three families,
\begin{equation}
\text{(A)}\ \zeta_\mu=\frac{c_1}{q^2}a^2\partial_\mu\sigma,\qquad \text{(B)}\ \zeta_\mu=\frac{c_2}{q}a^2\sigma\,\delta^\eta_\mu,\qquad \text{(C)}\ \zeta_\mu=\frac{c_3}{q}a^2\sigma\,\delta^z_\mu,
\end{equation}
each normalised so that $\mathcal L_\zeta\bar g=\mathcal O(h)$ uniformly in time. In the normalisation of Appendix~\ref{app:second} the change of $\langle E^{(2)}\rangle$ is of order $X^0$:
\begin{center}
\begin{tabular}{lcccc}
\toprule
 & $(\eta\eta)$ & $(\eta z)$ & $(zz)$ & $(xx)=(yy)$\\
\midrule
A & $-2H^2c_1(18H^2K-f_R)$ & $0$ & $2H^2c_1(6H^2K-f_R)$ & $12H^4c_1K$\\
B & $-\tfrac12H^2c_2^2(18H^2K-f_R)$ & $0$ & $\tfrac12H^2c_2^2(6H^2K-f_R)$ & $3H^4c_2^2K$\\
C & $\tfrac12H^2c_3^2(18H^2K-f_R)$ & $0$ & $-\tfrac12H^2c_3^2(6H^2K-f_R)$ & $-3H^4c_3^2K$\\
\bottomrule
\end{tabular}
\end{center}
This is $(H/k)^2$ times Eq.~(\ref{eq:main}). The factor $a^2$ in family A matters: without it $\mathcal L_\zeta\bar g/h$ grows like $X^2$, and one finds an apparent $X^2$ change proportional to $c_1(H/q)^2$ that is purely an artefact of the normalisation.

\paragraph{Wave packets.}
The same argument applies to the two-mode kernel of Appendix~\ref{app:packets}. With $\sigma$ replaced by the polarised pair $S_1e^{iq_1z}$, $S_2e^{-iq_2z}$ in the gauge vectors (A)--(C), with $q$ replaced by the carrier wavenumber $(q_1+q_2)/2$, the change of the kernel is $-E^{(1)}$ acting on $X_{12}(\eta)\,e^{i(q_1-q_2)z}$. Expanded in $e=(q_1-q_2)/q$, its order-$e^0$ part reproduces the table above, and its order-$e$ part grows at most like $X$ in all four components and for all three families. The physical order-$e$ term, $-i\Delta X^3\hat k_\mu\hat k_\nu$, is therefore gauge invariant up to corrections of relative order $(H/k)^2$.

\section{Second implementation and an exact case}\label{app:conformal}
For $g=e^{2w}\eta_{\mu\nu}$, with $w_a=\partial_aw$ and $\Box_\eta$ and contractions taken with $\eta_{\mu\nu}$,
\begin{gather}
\Gamma^c{}_{ab}=\delta^c_aw_b+\delta^c_bw_a-\eta_{ab}\eta^{cd}w_d,\nonumber\\
R_{ab}=-2(w_{ab}-w_aw_b)-\eta_{ab}\big(\Box_\eta w+2(\partial w)^2\big),\nonumber\\
R=-6e^{-2w}\big(\Box_\eta w+(\partial w)^2\big),\nonumber\\
R_{abcd}=g_{ac}S_{bd}+g_{bd}S_{ac}-g_{ad}S_{bc}-g_{bc}S_{ad},\qquad S_{ab}=\tfrac12\big(R_{ab}-\tfrac R6g_{ab}\big),\nonumber\\
G=\tfrac23R^2-2R_{ab}R^{ab}.
\end{gather}
Leaving $s_P$ in Eq.~(\ref{eq:P}) free, the Bianchi identity at orders $\epsilon$ and $\epsilon^2$ for arbitrary $\sigma$ has the unique solution $s_P=-1$. The off-diagonal components $(\eta x)$, $(xy)$, $(xz)$ vanish identically, and $E^{(0,1,2)}$ agree identically with the covariant implementation, off shell and for arbitrary $\sigma(\eta,z)$.

For $f_R=18H^2K$ one has $m^2=2H^2$ and the mode $s=(H/q)Xe^{iX}$ is exact. The averaged source is then, with no truncation,
\begin{align}
\langle E^{(2)}_{\eta\eta}\rangle={}&-18H^4\big[(16H^4f_{GG}+16H^2f_{RG}+3f_{RR})X^2+3W\big],\nonumber\\
\langle E^{(2)}_{\eta z}\rangle={}&54H^4K\,X^2,\nonumber\\
\langle E^{(2)}_{zz}\rangle={}&18H^4\big[-(80H^4f_{GG}+32H^2f_{RG}+3f_{RR})X^2+W\big],\nonumber\\
\langle E^{(2)}_{xx}\rangle={}&18H^4\big[(32H^4f_{GG}+8H^2f_{RG})X^2+W\big],
\end{align}
with
\begin{align}
W={}&3456H^8f_{GGG}+2592H^6f_{RGG}+648H^4f_{RRG}+54H^2f_{RRR}\nonumber\\
&+240H^4f_{GG}+72H^2f_{RG}+3f_{RR}.
\end{align}
At $f_R=18H^2K$ one has $\Delta=144H^6(f_{RG}+4H^2f_{GG})$, and the $X^2$ coefficients above agree with those of Appendix~\ref{app:second}.

\section{Direct variation of the action}\label{app:action}
Take $g_0=a^2e^{2\epsilon\chi}\eta_{\mu\nu}$ with arbitrary $\chi(\eta,z)$ and expand $f$ to third order about de~Sitter. The term $f_G\,G$ can be dropped, since $\sqrt{-g}\,G$ is a total derivative in four dimensions. For the variations $\delta g_{\eta\eta}=2\lambda a_1g_{0\eta\eta}$, $\delta g_{xx,yy}=2\lambda b_1g_{0xx,yy}$, $\delta g_{zz}=2\lambda c_1g_{0zz}$ and $\delta g_{\eta z}=\lambda d_1$, the Euler--Lagrange derivatives of the $\mathcal O(\lambda)$ Lagrangian obey
\begin{gather}
\frac{\delta S}{\delta a_1}=-2\sqrt{-g}E^\eta{}_\eta,\qquad
\frac{\delta S}{\delta b_1}=-2\sqrt{-g}\big(E^x{}_x+E^y{}_y\big),\nonumber\\
\frac{\delta S}{\delta c_1}=-2\sqrt{-g}E^z{}_z,\qquad
\frac{\delta S}{\delta d_1}=-2\sqrt{-g}E^{\eta z},
\end{gather}
identically at orders $\epsilon^0$, $\epsilon^1$, $\epsilon^2$ and off shell, with $E$ from Eq.~(\ref{eq:E}).

\section{Wave packets}\label{app:packets}
Write $E^{(2)}[\sigma]=B(\sigma,\sigma)$ and define $\mathcal K_{\mu\nu}(\eta;q_1,q_2)=\tfrac12B_{\mu\nu}(S_1e^{iq_1z},S_2e^{-iq_2z})$, where $S_1$ is the mode of wavenumber $q_1$ and $S_2$ the complex conjugate of the mode of wavenumber $q_2$. Using both mode equations one finds $\bar\nabla^\mu\big(\mathcal K_{\mu\nu}e^{i(q_1-q_2)z}\big)=0$ identically in $q_1,q_2$. We factor out $e^{i(q_1-q_2)(z-\eta)}$, set $q_{1,2}=q(1\pm e/2)$ and expand in $e$. At leading order in $H/k$ the $e^0$ term is the single-mode result of Appendix~\ref{app:second}; the $e^1$ term consists of the single structure $-i\Delta X^3\hat k_\mu\hat k_\nu$, which is absent in $f(R)$, with no $X^3$ term in $(xx)$ or $(yy)$; and the $e^2$ term is at most of order $X^2$. Since $-ieX=i(q_1-q_2)\eta$ and $i(q_1-q_2)$ acts as $\partial_u$ on the envelope, the order-$e$ term equals $\Delta X^2\,\eta\partial_u\,\hat k_\mu\hat k_\nu$ acting on the envelope. Adding the $e^0$ term and replacing $X^2$ by $2\langle\sigma_{,z}^2\rangle/H^2$ as in Appendix~\ref{app:second},
\begin{equation}
\kappa^2t_{\mu\nu}=6f_R\langle\sigma_{,z}^2\rangle\hat k_\mu\hat k_\nu-\frac{2\Delta}{H^2}\Big[\langle\sigma_{,z}^2\rangle\big(\hat t\hat t+\hat x\hat x+\hat y\hat y-\hat z\hat z\big)_{\mu\nu}+\eta\,\partial_u\langle\sigma_{,z}^2\rangle\,\hat k_\mu\hat k_\nu\Big],
\end{equation}
to first order in envelope gradients, which is Eq.~(\ref{eq:packet}). As a cross-check in position space, take an arbitrary profile $F(u)=\langle\sigma_{,z}^2\rangle$ and an undetermined coefficient $c$ in front of the gradient term. Then
\begin{equation}
\bar\nabla^\mu t_{\mu\eta}\propto-\Delta H^2\eta^4q^2\,(c-1)\,F'(u),\qquad \bar\nabla^\mu t_{\mu z}\propto\Delta H^2\eta^4q^2\,(c-1)\,F'(u),
\end{equation}
so conservation requires $c=1$.

\section{Off de Sitter: analytic relations for $f=R+\alpha(R+\lambda G)^2$}\label{app:offdS}
Here $\phi=1+2\alpha\chi$, $\xi=2\alpha\lambda\chi$ and $V=\alpha\chi^2$ with $\chi=R+\lambda G$; Eq.~(\ref{eq:E}) is used without truncation.

\paragraph{Background.}
On flat FLRW, in cosmic time,
\begin{align}
\tfrac13E_{00}={}&H^2+36\alpha H^2\dot H+12\alpha H\ddot H-6\alpha\dot H^2\nonumber\\
&+\lambda\alpha\big(-48H^6+288H^4\dot H+96H^3\ddot H+48H^2\dot H^2\big)\nonumber\\
&+\lambda^2\alpha\big(-96H^8+576H^6\dot H+192H^5\ddot H+288H^4\dot H^2\big).
\end{align}
At $\lambda=0$ this is the Starobinsky equation. Setting $\dot H=-\eH H^2$ and dropping $\ddot H$ and $\dot H^2$, which are $\mathcal O(\eH^2)$, gives $\eH=D/(18H^2K)$ with $D=1-48\alpha H^2u(1+2u)$ and $K=2\alpha(1+4u)^2$.

\paragraph{Growing mode.}
With $\alpha=1/[48u(1+2u)]$ and units $H_{\rm dS}=1$, $H=1+\delta e^{st}$ solves both background equations at $\mathcal O(\delta)$ if and only if $s^2+3s+m^2=0$.

\paragraph{Linear scalar sector.}
In Newtonian gauge, $g=a^2[-(1+2\Phi)d\eta^2+(1-2\Psi)d\bm x^2]$, on an arbitrary FLRW background, we write $\Psi={\rm Re}[(s_0+\dots)e^{i\theta}e^{iqz}]$, $\Phi={\rm Re}[(p_0+\dots)e^{i\theta}e^{iqz}]$ and $\theta'=-(qc_0+c_1+\dots)$. At order $q^4$ every component is proportional to
\begin{equation}
L=(1+4u)p_0+s_0\big[3c_0^2(1+4u)-2-8u-8\lambda\dot H\big],
\end{equation}
with prefactors $4\alpha(1+4u)\{1,-c_0,c_0^2\}/a^2$ for $(\eta\eta)$, $(\eta z)$, $(zz)$ and $4\alpha[(c_0^2-1)(1+4u)-4\lambda\dot H]/a^2$ for $(xx)$, so $L=0$. At order $q^3$ the equations are satisfied identically. At order $q^2$ the $(xx)$ component fixes
\begin{equation}
c_s^2-1=\frac{16\lambda\,\mathcal N}{3(1+4\lambda H^2)\,\mathcal M},
\end{equation}
where
\begin{align}
\mathcal N={}&\dot H+36\alpha H^2\dot H-9\alpha H\ddot H-3\alpha\dddot H\nonumber\\
&+\lambda\alpha\big(144H^4\dot H-72H^3\ddot H+72H^2\dot H^2-24H^2\dddot H-24H\dot H\ddot H-24\dot H^3\big)\nonumber\\
&+\lambda^2\alpha\big(192H^6\dot H-144H^5\ddot H+288H^4\dot H^2-48H^4\dddot H\nonumber\\
&\qquad-96H^3\dot H\ddot H+288H^2\dot H^3\big),\nonumber\\
\mathcal M={}&1+24\alpha H^2+12\alpha\dot H+\lambda\alpha\big(48H^4+240H^2\dot H+48H\ddot H\big)\nonumber\\
&+\lambda^2\alpha\big(768H^4\dot H+192H^3\ddot H+384H^2\dot H^2\big).
\end{align}
Hence $c_s^2=1$ on de~Sitter and in $f(R)$. In slow roll ($\ddot H=\dddot H=0$, $\dot H=-\eH H^2$), using the background relation above, $c_s^2=1-8\eH u(24u^2+16u+3)/[3(2u+1)^2(4u+1)]$.

\paragraph{Fluid-dominated backgrounds.}
The expression for $c_s^2$ above holds on any FLRW background. For a fluid with constant $w$ we set $\dot H=-\tfrac32(1+w)H^2$, $\ddot H=\tfrac92(1+w)^2H^3$ and $\dddot H=-\tfrac{81}{4}(1+w)^3H^4$, and expand in $\alpha H^2$. The leading term is
\begin{equation}
c_s^2=1-\frac{8u(1+w)}{1+4u},
\end{equation}
which tends to $-1-2w$ as $u\to\infty$ and vanishes at $u=1/4$ for $w=0$ and at $u=3/20$ for $w=1/3$. The first correction is
\begin{equation}
12\alpha H^2\,u(1+w)\big[(360w^2+528w+136)u+27w^2+24w-11\big].
\end{equation}

\paragraph{Tensor speed.}
For $h_{xy}=a^2\gamma(\eta)e^{iqz}$ on an arbitrary FLRW background the linear $(xy)$ equation is of second order, and the ratio of its $q^2\gamma$ and $\gamma''$ coefficients is identically $c_T^2=(F+4\ddot\xi)/(F+4H\dot\xi)$ of Ref.~\cite{DGS}, both for $f=R+\alpha(R+\lambda G)^2$ and for $f=R+\alpha R^2+\gamma G^2$.

\section{Non-degenerate class and numerical evaluations}\label{app:numerics}
The statements below can be reproduced with the deposited code; the file names are given in brackets.

\emph{Non-degenerate class} [\texttt{dispersion\_screen.py}]. For $f=R+\alpha R^2+\gamma G^2$ on the slow-roll background $R=6(2H^2+\dot H)$, $G=24H^2(H^2+\dot H)$ with $\dot H=-\eH H^2$, the coefficient $B_2$ of Eq.~(\ref{eq:B2}) is, to order $\eH^2$, with $a\equiv\alpha H^2$ and $g\equiv|\gamma|H^6$,
\begin{equation}
H^2B_2=\mp\frac{128\,a\,g\,\eH^2}{3(a\pm16g)(24a+1)}\qquad(\gamma\gtrless0).
\end{equation}
For $\gamma>0$ the instability rate $\sqrt{|B_2|}k_{\rm phys}^2$ exceeds $H$ for
\begin{equation}
\frac{k_{\rm phys}}{H}>\Big[\frac{3(a+16g)(24a+1)}{128\,a\,g\,\eH^2}\Big]^{1/4}.
\end{equation}
For $\gamma<0$ and $a>16g$ the group velocity $d\omega/dk$ exceeds unity for
\begin{equation}
\frac{k_{\rm phys}}{H}>\frac{1}{32\,\eH}\sqrt{\frac{(144a+6)(a-16g)}{a\,g}} .
\end{equation}
The growth rate $\sqrt{|B_2|}k_{\rm phys}^2$ exceeds $H$ for $k_{\rm phys}>19.6H$, $28.4H$, $17.8H$ and $119H$ at $(\alpha H^2,\gamma H^6,\eH)=(10,1,10^{-2})$, $(10^4,10^2,10^{-2})$, $(1,1,10^{-2})$ and $(10^9,10^4,1/60)$.

\emph{Growing-mode background} [\texttt{quasi\_ds.py}]. For each $u$ below, the WKB solution of the linear equations holds through $\mathcal O(q^0)$ with all residuals vanishing, the $q^4$ and $q^3$ terms of the averaged second-order source vanish, and its $q^2$ term at $\delta^0$ equals Eq.~(\ref{eq:main}). The table gives the growth rate $s$ of the homogeneous mode, the scalar sound speed, and the remainders of the $\mathcal O(\eH)$ part after the local-$H$ value of Eq.~(\ref{eq:main}) has been subtracted, divided by the de~Sitter value of each component, in units of $\eH$.
\begin{center}
\begin{tabular}{ccccccc}
\toprule
$u$ & $s/H_{\rm dS}$ & $(c_s^2-1)/\eH$ & $(\eta\eta)$ & $(\eta z)$ & $(zz)$ & $(xx)$\\
\midrule
$1/200$ & $0.0258$ & $-2/51$ & $-0.59$ & $-0.56$ & $-0.57$ & $1.22$\\
$1/100$ & $0.0500$ & $-1/13$ & $-0.69$ & $-0.62$ & $-0.64$ & $1.10$\\
$1/50$ & $0.0940$ & $-4/27$ & $-0.87$ & $-0.74$ & $-0.78$ & $0.88$\\
$1/20$ & $0.1997$ & $-1/3$ & $-1.43$ & $-1.08$ & $-1.20$ & $0.28$\\
$1/10$ & $0.3197$ & $-4/7$ & $-2.31$ & $-1.61$ & $-1.84$ & $-0.58$\\
$1/5$ & $0.4571$ & $-8/9$ & $-3.92$ & $-2.54$ & $-2.92$ & $-1.94$\\
$1/2$ & $0.6148$ & $-4/3$ & $-7.50$ & $-4.50$ & $-5.13$ & $-4.47$\\
$1$ & $0.6932$ & $-8/5$ & $-10.97$ & $-6.29$ & $-7.07$ & $-6.58$\\
\bottomrule
\end{tabular}
\end{center}

\emph{Tensor speed on the slices $u=1/50$ and $1/20$} [\texttt{tensor\_speed.py}]. $c_T^2-1=\tfrac4{27}\eH$ and $\tfrac13\eH$.

\bibliography{frg_isaacson}
\end{document}